\documentclass[aps,showpacs,superscriptaddress,nofootinbib,amsmath,amssymb]{revtex4}
 \usepackage{graphicx}
\usepackage{dcolumn}
\usepackage{bm}
\usepackage{lipsum}
\usepackage{adjustbox}
\usepackage{multirow}
\usepackage{amsmath}
\usepackage{amssymb}
\usepackage{enumerate}

\graphicspath{./fig}
\begin{document}
 \title{Nonperturbative and higher order perturbative effects in deep inelastic $\nu_\tau/\bar\nu_\tau-$nucleon scattering}
\author{V. Ansari}
\affiliation{Department of Physics, Aligarh Muslim University, Aligarh - 202002, India}
\author{M. Sajjad Athar}
\author{H. Haider}
\affiliation{Department of Physics, Aligarh Muslim University, Aligarh - 202002, India}
\author{S. K. Singh}
\affiliation{Department of Physics, Aligarh Muslim University, Aligarh - 202002, India}
\author{F. Zaidi\footnote{Corresponding author: zaidi.physics@gmail.com}}
\affiliation{Department of Physics, Aligarh Muslim University, Aligarh - 202002, India}


\begin{abstract}
 The effect of nonperturbative and higher order perturbative corrections to all the free nucleon structure functions ($F_{iN}(x,Q^2); i=1-5$) in the DIS of $\nu_\tau/{\bar\nu}_\tau$ on nucleon is studied. 
 The target mass correction (TMC) and higher twist (HT) effects are incorporated following the works of Kretzer et al. and Dasgupta et al., respectively. 
 The evaluation of the nucleon structure functions has been performed by using the MMHT 2014 parameterization of the parton distribution functions (PDFs). The calculations have been performed at the next-to-leading(NLO) order.
 These nucleon structure functions (SF) are used to calculate the DIS cross section by further including the kinematical corrections due to $\tau$-lepton mass. 
  Due to the inclusion of lepton mass two additional structure functions $F_{4N}(x,Q^2)$ and $F_{5N}(x,Q^2)$ become non-negligible.
 The results for the nucleon structure functions, differential and total scattering cross sections are presented. The various effects considered in this work are effective in the different regions of $x$ and $Q^2$, and quite important in the energy region of 
 $E_{\nu_\tau/{\bar\nu}_\tau} < 15$ GeV. A comparative study of our results with the existing
 results in the literature for the cross 
 sections is made in the energy region of interest for the DUNE, SHiP, DsTau
 and HyperK experiments proposed to be done in the near future.
  \end{abstract}
\maketitle
   \section{Introduction} 
  With the discovery of $\tau$-lepton in 1975 by the SLAC-LBL (Stanford Linear Accelerator Center-Lawrence Berkeley Laboratory) collaboration~\cite{Perl:1975bf}, predictions were made for the
  existence of its weak isospin neutral partner $\nu_\tau$. The idea of three flavors of neutrinos received very strong support in 1989 
  when LEP (Large Electron-Positron Collider) concluded the presence of three active neutrinos~\cite{Decamp:1989tu,Aarnio:1989tv,Adeva:1989mn}. 
 The $\nu_\tau$s were first observed 
  by the DONuT (Direct Observation of the Nu Tau) Collaboration~\cite{Kodama:2007aa}, where the energy dependent cross section was reported to be $0.39 \pm 0.13 \pm 0.13 \times 10^{-38}$ cm$^2/$GeV, against the Standard Model prediction of
  $0.5 \times 10^{-38}$ cm$^2/$GeV~\cite{Kodama:2000mp}. In an accelerator experiment $\nu_\tau$s were observed in the $\nu_\mu \rightarrow \nu_\tau$ appearance mode by the OPERA (Oscillation Project with Emulsion-tRacking Apparatus) Collaboration~\cite{Agafonova:2018auq, Agafonova:2015jxn, Agafonova:2014ptn, Agafonova:2014bcr, Agafonova:2013dtp} at CERN. 
  In total there are fourteen $\nu_\tau + \bar\nu_\tau$ events reported by these two collaborations and these are the only two experiments 
  which have reported a direct observation of tau neutrinos  through the charged current interactions. In the atmospheric neutrino experiment 
  at Super-Kamiokande (SK), $\nu_\tau$ events were statistically inferred from the multi-GeV atmospheric neutrinos. The Super-Kamiokande collaboration has 
  also reported $\nu_\tau$ charged current cross section averaged over $\nu_\tau$ flux in the energy range 3.5 GeV to 70 GeV to be $0.94 \pm 0.20 \times 10^{-38}$ cm$^2$ against the  Standard Model prediction of 
  $0.64 \pm 0.20 \times 10^{-38}$ cm$^2$~\cite{Abe:2012jj, Li:2017dbe}. Recently $\nu_\tau$ events have been reported by the 
  IceCube Collaboration~\cite{Aartsen:2019tjl} in the DeepCore subarray of the observatory using atmospheric neutrinos in 
  the energy range 5.6 GeV to 56 GeV.

It has been realized that to test the Standard Model predictions and check the validity of the Lepton Universality 
hypothesis; the interaction cross sections for all the three flavors of neutrinos should be known to high accuracy requiring better measurements of the $\nu_\tau/{\bar\nu}_\tau$-nucleon cross sections. 
Furthermore, $\nu_\tau$ interaction studies are also required to better determine the properties of the third neutrino weak eigenstate to have precise understanding of the neutrino oscillation 
parameters.
Keeping this in mind new proposed experiments are coming up, for example, 
the SHiP (Search for Hidden Particles)~\cite{SHiP:2018xqw, DiCrescenzo:2016irr} and the DsTau~\cite{Aoki:2019jry} experiments. The SHiP experiment at CERN (European Council for Nuclear Research) is planning to measure $\nu_\tau/{\bar\nu}_\tau$ events using proton-proton collisions giving 
rise to $D_s$-mesons which subsequently decay to $\tau \nu_\tau$ and the plan is to study $\nu_\tau/{\bar\nu}_\tau$
scattering on lead target with a few hundred times larger statistics than observed in the DONuT experiment. 
They also plan to measure weak $F_4$ and $F_5$ structure functions. The DsTau Collaboration plans to study $\nu_\tau/{\bar\nu}_\tau$ 
flux for neutrino accelerator experiment by using tau leptonic decay of $D_s$ mesons produced in 400 GeV proton interactions and the goal
is to reduce the $\nu_\tau/{\bar\nu}_\tau$ flux uncertainty to 10$\%$ level for reducing the systematics in the future neutrino 
beam experiment looking for tau events in the charged current (CC) induced $\nu_\tau/\bar\nu_\tau$ interaction.

Moreover, in the neutrino oscillation physics sector, the DUNE (Deep Underground Neutrino Experiment)~\cite{Abi:2020qib, Abi:2018dnh, Abi:2020mwi} and T2HK(Tokai to Hyper-Kamiokande)~\cite{Abe:2018uyc,Yu:2018atd}  are the proposed 
precision measurement experiments for studying the oscillation parameters which will be capable of explaining CP violation in the lepton sector. These experiments would also be sensitive to $\nu_\tau/{\bar\nu}_\tau$ detection, i.e. these experiments would also have the ability to test the validity of three neutrino hypothesis and if possible some new physics 
associated with neutrinos. 

DUNE would be using 40kT liquid argon detector known as LArTPC(Liquid Argon Time Projection Chamber) for the long baseline at a distance of about
1300km from the Fermilab. The neutrino beam intensity would be high($10^{21}$POT) and it would have exquisite track reconstruction having a tracking resolution of around several millimeters. 
The flux averaged muon neutrino energy($<E_{\nu_\mu}>$) is expected to be around 4  GeV, and the beam energy peak around 2.5 GeV with a broad range of 
neutrino energies almost extending up to 20 GeV. At these energies the contribution to the cross section would come from the quasielastic, 
resonance excitations as well as the deep inelastic scattering processes. It is expected that $\nu_\tau/{\bar\nu}_\tau$ events in the appearance mode at 
DUNE would be between 100 to 1000 and the observations would be made through the charged current interactions producing $\tau$-leptons 
which would subsequently decay and be identified by their decay products. 
The decay length of $\tau$-leptons produced in the charged current $\nu_\tau/{\bar\nu}_\tau$ interaction would be significantly smaller than the resolution of the DUNE detector implying that one would have to reconstruct the $\tau$-decay production in order to classify 
the incoming neutrino as $\nu_\tau$.  

 Theoretically $\nu_\tau/{\bar\nu}_\tau - N$ DIS calculations have been performed by various groups like 
 Kretzer and Reno~\cite{Kretzer:2002fr, Kretzer:2003iu}, Jeong and Reno~\cite{Jeong:2010nt}, Hagiwara et al.~\cite{Hagiwara:2003di}, Paschos and Yu~\cite{Paschos:2001np}, Gazizov et al.~\cite{Gazizov:2016} on free nucleons. The $\nu_\tau/{\bar\nu}_\tau-$nucleon$(N)$ cross sections have large uncertainties as compared to 
$\nu_e/{\bar\nu}_e - N$  and $\nu_\mu/{\bar\nu}_\mu - N$ cross sections~\cite{Conrad:2010mh}. As observed by Conrad et al. ~\cite{Conrad:2010mh}, the present studies on the charged current $\nu_\tau/{\bar\nu}_\tau-N$ cross sections have large variations  arising due to the uncertainties in the nucleon structure functions. For example, the flux averaged total charged current $\nu_\tau + {\bar\nu}_\tau$ cross section ($<\sigma>$) in the energy range $6 < E< 30$ GeV varies between 0.3 to 0.58 ($\times 10^{-38}$ cm$^2$)
 ~\cite{Hagiwara:2003di, Paschos:2001np, Kretzer:2002fr, Naples:2003ne, Casper:2002sd}, or the expected number of CC $\nu_\tau/\bar\nu_\tau$ events/100 kT.yr on an isoscalar target range between $\sim$52 to $\sim$95 events, which is a factor of almost 2. 
 In this energy range a major contribution is expected from the DIS process for $E > 10$ GeV. Moreover, these interactions would be studied in a nuclear target like $^{40}Ar$ in the case of DUNE, and $^{208}Pb$ in the case of CERN experiments, where nuclear medium effects become important.
 It is therefore important to study the various uncertainties in the $\nu_\tau/{\bar\nu}_\tau -N$  cross sections on free nucleons as well as 
 nuclear medium effects in medium and heavy nuclei in a wide range of energy transfer ($\nu=E_\nu -E_l$) and the four momentum transfer  square($Q^2 \ge 0$) to the target. In this work, we focus on the various uncertainties involved in the theoretical calculation of $\nu_\tau/\bar\nu_\tau-$nucleon cross sections in the deep inelastic scattering region. This is because the nucleon structure functions are the basic inputs in the determination of the scattering cross section, and a good  understanding of the nucleon structure functions become quite important. In the region of low and moderate $Q^2$, the perturbative and nonperturbative 
QCD corrections such as $Q^2$ evolution of parton distribution functions from leading order (LO) to higher order terms (next-to- leading order (NLO), 
next-next-to-leading order (NNLO), ...), the effects of target mass corrections (TMC) due to the massive quarks production (e.g. charm, bottom, top) and 
higher twist (twist-4, twist-6, ...) because of the multiparton correlations, become important. The nonperturbative effects are specifically 
important in the kinematical region of high $x$ and low $Q^2$, sensitive to some of the oscillation parameters, and therefore it is of considerable 
experimental interest to the long baseline oscillation experiments. These effects have not been discussed extensively in the context of 
the deep inelastic $\nu_\tau/{\bar\nu}_\tau - N$ scattering.

In this work, we have evaluated the nucleon structure functions by using the MMHT parameterization for the parton distribution 
functions~\cite{Harland-Lang:2014zoa} up to 
next-to-leading order in the four flavor($u~ d,~ s,$ and $c$) scheme following Ref.~\cite{Kretzer:2003iu}.  The QCD corrections  due to the nonperturbative higher 
twist effect is incorporated by using the approach discussed in Ref.~\cite{Dasgupta:1996hh} and the target mass correction is included following 
the works of Kretzer et al.~\cite{Kretzer:2003iu} to calculate all the 
structure functions $F_{iN}(x,Q^2);~~(i=1-5)$ as a function of $x$ for various $Q^2$. While performing the calculations for the free nucleon case, the QCD corrections have been incorporated and then differential and total cross sections are calculated including the kinematic factors involving $\tau$ lepton mass. The uncertainties in the cross sections arising due to the use of different approaches for NLO evolution of parton densities such as given by Kretzer et al.~\cite{Kretzer:2002fr}, and Vermaseren et al.~\cite{Vermaseren:2005qc} and Moch et al.~\cite{Moch:2004xu, Moch:2008fj} have been discussed. The results of the cross sections are compared with the earlier results available in the  literature like that of Kretzer et al.~\cite{Kretzer:2002fr, Kretzer:2003iu}, Jeong et al.~\cite{Jeong:2010nt}, Anelli et al.~\cite{Anelli:2015pba}, Paschos et al.~\cite{Paschos:2001np}, Hagiwara et al.~\cite{Hagiwara:2003di} and Gazizov et al.~\cite{Gazizov:2016} some of which have been used in the experimental analysis to obtain $\nu_\tau-$nucleon cross sections by the DONuT~\cite{Kodama:2007aa, Kodama:2000mp}, SuperK~\cite{Abe:2012jj, Li:2017dbe} and IceCube~\cite{Aartsen:2019tjl} collaborations as well as in the simulation studies made for the proposed experiments by SHiP~\cite{SHiP:2018xqw, DiCrescenzo:2016irr}, DsTau~\cite{Aoki:2019jry} and DUNE~\cite{Abi:2018dnh, Abi:2020mwi, Abi:2020qib} collaborations. 

An important aspect of our present study is to focus on the importance of the kinematic cuts on $Q^2$ and $W$ which have also been used in various simulation studies while applying the DIS formalism. The kinematic cut has varied in literatures from 1.4 GeV to 2.0 GeV in the case of $W$ while it has been extrapolated to lower values of $Q^2 (< 1 ~$ GeV$^2)$ in some of the works~\cite{Hagiwara:2003di, Kretzer:2002fr,Jeong:2010nt}. In this work, we have performed  all the numerical calculations for 
 $Q^2 > 1 $GeV$^2$. We have also presented and discussed in some detail the effect of $\tau$ lepton mass on the total and differential cross sections arising due to the kinematic corrections in the contribution of $F_{1N}(x,Q^2)$, $F_{2N}(x,Q^2)$ and $F_{3N}(x,Q^2)$ as well as  due to the presence of additional structure
functions $F_{4N}(x,Q^2)$ and $F_{5N}(x,Q^2)$ by comparing the ${\sigma_\tau}$ and ${\sigma_\mu}$ cross sections for various energies in the DIS region of $\nu_\tau/\bar\nu_\tau-N$ scattering. The present work is assumed to be an update on the earlier work on the $\nu_\tau-$nucleon scattering in the DIS region in the presence of nonperturbative QCD effects focusing on the importance of kinematic cuts on $W$ and $Q^2$, and is extension of our work performed recently for the $\nu_\mu/{\bar\nu}_\mu - N$~\cite{Zaidi:2019asc} and 
$l^\pm - N$~\cite{Zaidi:2019mfd} DIS studies. This paper is organized as follows. 

In section~\ref{sec_formalism}, we present in brief the formalism for calculating the nucleon structure functions and cross sections for the  $\nu_\tau/\bar\nu_\tau-$nucleon scattering in the deep inelastic region
including the corrections due to kinematical TMC and dynamical twist-4 effects. 
In section~\ref{results}, numerical results for the structure functions and cross sections are presented and the role of kinematic cuts on the cross sections are discussed. In section~\ref{summary} we summarize our results with conclusions. 

\begin{figure}[h]
\begin{center}
 \includegraphics[height=4.0 cm, width=6 cm]{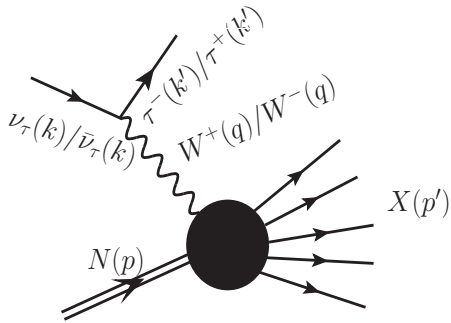}
  \end{center}
 \caption{Feynman representation for the $\nu_\tau/\bar\nu_\tau$ induced DIS process off free nucleon target.}
 \label{lag_weak}
\end{figure}
 \section{Formalism: $\nu_\tau-N$ DIS process}\label{sec_formalism}
 The basic reaction for the (anti)neutrino induced charged current deep inelastic scattering process on a free nucleon target is given by
\begin{eqnarray}\label{reaction}
 \nu_\tau(k) / \bar\nu_\tau(k) + N(p) \rightarrow \tau^-(k') / \tau^+(k') + X(p'),
\end{eqnarray}
where $k$ and $k'$ are the four momenta of incoming and outgoing lepton, $p$ and $p'$ are the four momenta of the target nucleon and the jet of
hadrons produced 
in the final state, respectively. 
This process is mediated by the $W$-boson ($W^\pm$) and the invariant matrix element corresponding to the reaction given in Eq.\ref{reaction} is written as
\begin{equation}\label{matrix}
 -i{\cal M}=\frac{iG_F}{\sqrt{2}}\;l_\mu \;\left(\frac{M_W^2}{q^2-M_W^2} \right)\;\langle X|J^\mu|N\rangle\;,
\end{equation}
where $G_F$ is the Fermi coupling constant, $M_W$ is the mass of $W$ boson, and $q^2=(k-k')^2$ is the four momentum transfer square. $l_\mu$ is the leptonic 
current and $\langle X|J^\mu|N\rangle$ is the 
hadronic current for the neutrino induced reaction (shown in Fig.~\ref{lag_weak}). 

\begin{figure}[h]
\begin{center}
\includegraphics[height=4.5 cm, width=8.5 cm]{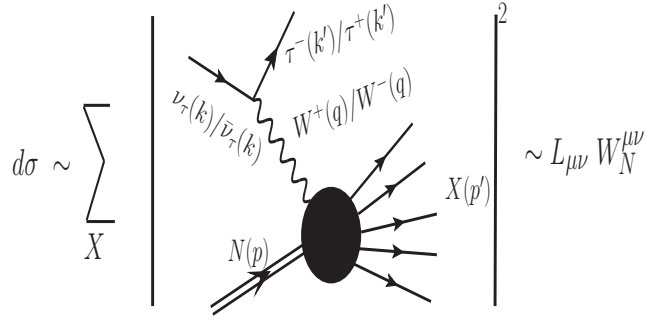}
\end{center}
\caption{ $\nu_\tau({\bar\nu}_\tau) - N$ inclusive scattering where the 
 summation sign represents the sum 
over all the hadronic states such that the cross section($d\sigma$) for the deep inelastic scattering  
$\propto L_{\mu \nu} W_{N}^{\mu \nu}$.}
\label{fg:fig1}
\end{figure}

The general expression of double differential scattering cross section (DCX) corresponding to the reaction given in Eq.~\ref{reaction} (depicted in Fig.~\ref{fg:fig1}) in the laboratory frame is expressed as:
\begin{equation}
\label{eq:w1w2w3}
\frac{ d^2\sigma  }{ dx dy } =  \frac{y M_N}{\pi }~\frac{E}{E'}~\frac{|{\bf k^\prime}|}{|{ \bf k}|}\;   \bar\sum \sum |{\cal M}|^2 \;,
\end{equation}
where $x$ and $y$ are the scaling variables which lie in the following ranges:
\begin{equation}
\left.
\begin{array}{l}
 \frac{m_l^2}{2M_N (E - m_l)} \le x \le 1\\
 a-b \le y\le a+b,
\end{array}
\right\}
\end{equation}
with
\begin{eqnarray}
 a&=&\frac{1-m_l^2\Big(\frac{1}{2M_NE x}+\frac{1}{2E^2} \Big)}{2\Big(1+\frac{M_N x}{2E}\Big)}\\
 b=&=&\frac{\sqrt{\left(1-\frac{m_l^2}{2 M_N E x}\right)^2-\frac{m_l^2}{E^2}}}{2\Big(1+\frac{M_N x}{2E}\Big)}.
\end{eqnarray}
$\nu=E-E'$ is the energy transfer, $E(E')$ is the energy of the incoming(outgoing) lepton, $M_N$ is the nucleon mass, $m_l$ is the charged lepton mass and $\bar\sum \sum |{\cal M}|^2$ is the 
invariant matrix element square which is given in terms of the leptonic ($L_{\mu\nu} $) and hadronic ($ W^{\mu\nu}_N$) tensors as
\begin{equation}\label{amp_wk}
 \bar\sum \sum |{\cal M}|^2 = \frac{G_F^2}{2}~\left(\frac{M_W^2}{Q^2+M_W^2}\right)^2 ~L_{\mu\nu} ~W^{\mu\nu}_N,
\end{equation}
with $Q^2=-q^2\ge 0$. $L_{\mu \nu} $ is given by 
\begin{eqnarray}\label{lep_weak}
L_{\mu \nu} &=&8(k_{\mu}k'_{\nu}+k_{\nu}k'_{\mu}
-k.k^\prime g_{\mu \nu}  \pm i \epsilon_{\mu \nu \rho \sigma} k^{\rho} 
k'^{\sigma})\,.
\end{eqnarray}
Here the antisymmetric term arises due to the contribution from the axial-vector components with +ve sign
for antineutrino and -ve sign for neutrino. 
The hadronic tensor $W_{N}^{\mu \nu}$ is written in terms of the weak nucleon structure functions $W_{iN} (\nu,Q^2)~(i=1-6)$ as
\begin{eqnarray}\label{ch2:had_ten_N}
W_{N}^{\mu \nu} &=&
\left( \frac{q^{\mu} q^{\nu}}{q^2} - g^{\mu \nu} \right) \;
W_{1N} (\nu, Q^2)
+ \frac{W_{2N} (\nu, Q^2)}{M_N^2}\left( p^{\mu} - \frac{p . q}{q^2} \; q^{\mu} \right)
 \nonumber\\
&\times&\left( p^{\nu} - \frac{p . q}{q^2} \; q^{\nu} \right)-\frac{i}{2M_N^2} \epsilon^{\mu \nu \rho \sigma} p_{ \rho} q_{\sigma}
W_{3N} (\nu, Q^2) + \frac{W_{4N} (\nu, Q^2)}{M_N^2} q^{\mu} q^{\nu}\nonumber\\
&&+\frac{W_{5N} (\nu, Q^2)}{M_N^2} (p^{\mu} q^{\nu} + q^{\mu} p^{\nu})
+ \frac{i}{M_N^2} (p^{\mu} q^{\nu} - q^{\mu} p^{\nu})
W_{6N} (\nu, Q^2)\,.
\end{eqnarray}
The contribution of the term with $W_{6N} (\nu, Q^2)$ vanishes when contracted with the leptonic tensor. When $Q^2$ and $\nu$ become large the structure functions $W_{iN}  (\nu,Q^2);~(i=1-5)$ are generally expressed in terms of the dimensionless nucleon structure functions $F_{iN}  (x),\;\;i=1 - 5$ as: 
 \begin{eqnarray}\label{ch2:relation}
 F_{1N}(x) &=& W_{1N}(\nu,Q^2) \nonumber\\
 F_{2N}(x) &=& \frac{Q^2}{2xM_N^2}W_{2N}(\nu,Q^2)\nonumber\\
 F_{3N}(x) &=& \frac{Q^2}{xM_N^2}W_{3N}(\nu,Q^2) \nonumber\\
 F_{4N}(x) &=& \frac{Q^2}{2M_N^2}W_{4N}(\nu,Q^2) \nonumber\\
 F_{5N}(x) &=& \frac{Q^2}{2xM_N^2}W_{5N}(\nu,Q^2) \nonumber
\end{eqnarray}
 
Now the hadronic tensor may be written in terms of dimensionless nucleon structure functions $F_{iN}(x,Q^2)\;\;(i=1-5)$ as:
\begin{eqnarray}\label{had_weak_red}
W_{N}^{\mu \nu} 
& = &- g_{\mu\nu} F_{1N}(x,Q^2) + \frac{p_{\mu}p_{\nu}}{{p_\cdot q}} F_{2N}(x,Q^2) - i \epsilon_{\mu\nu\rho\sigma} \frac{p^{\rho} q^{\sigma}}{2 p_1\cdot q} F_{3N}(x,Q^2) + \nonumber \\
          &   & \frac{q_{\mu} q_{\nu}}{p\cdot q} F_{4N}(x,Q^2) +
                (p_{\mu}q_{\nu} + p_{\nu}q_{\mu}) F_{5N}(x,Q^2).
\end{eqnarray}
The expression for the differential scattering cross section for the $\nu_\tau/{\bar\nu}_\tau - N$ scattering 
given in Eq.~\ref{eq:w1w2w3} is written by using Eqs.~\ref{lep_weak} and \ref{had_weak_red} as:
\begin{small}
\begin{eqnarray}
 \frac{d^2\sigma}{dxdy}&=&\frac{G_F^2M_NE_\nu}{\pi(1+\frac{Q^2}{M_W^2})^2}
 \Big\{\Big[y^2x+\frac{m_l^2 y}{2E_\nu M_N}\Big]F_{1N}(x,Q^2)+
 \Big[\Big(1-\frac{m_l^2}{4E_\nu^2}\Big)-\Big(1+\frac{M_Nx}{2E_\nu}\Big)y\Big]F_{2N}(x,Q^2)\nonumber\\
 &\pm& \Big[xy\Big(1-\frac{y}{2}\Big)-
 \frac{m_l^2 y}{4E_\nu M_N}\Big]F_{3N}(x,Q^2)
 +\frac{m_l^2(m_l^2+Q^2)}{4E_\nu^2M_N^2 x}F_{4N}(x,Q^2)-\frac{m_l^2}{E_\nu M_N}F_{5N}(x,Q^2)\Big\}.\;\;\;
\end{eqnarray}
\end{small}
In general, the dimensionless nucleon structure functions are derived in the quark-parton model assuming Bjorken scaling in which they are written in terms of the parton distribution functions $q_i(x)$ and $\bar q_i(x)$ at the leading order as
\begin{eqnarray}\label{parton_wk}
F_{2}(x)  &=& \sum_{i} x [q_i(x) +\bar q_i(x)] \;;\;
x F_3(x) =  \sum_i x [q_i(x) -\bar q_i(x)]\;;\;\;
F_4(x)=0.
\end{eqnarray} 
 For example, in the case of $\nu(\bar{\nu})$-proton scattering 
 above the charm production threshold, $F_{2,3}(x)$ are given by:
\begin{eqnarray}\label{eq:pdf1}
F_{2p}^{\nu }(x)& = & 2 x [d(x) + s(x) + \bar{u}(x) +\bar{c}(x)]\;;\;\; F_{2p}^{\bar\nu}(x)=  2 x [u(x) + c (x)+ \bar{d}(x) +\bar{s}(x)]\nonumber\\
x F_{3p}^{\nu}(x)& = & 2 x [d(x) + s (x)- \bar{u}(x) -\bar{c}(x)]\;;\;\;
x F_{3p}^{\bar\nu}(x)= 2 x [u(x) + c(x) - \bar{d}(x) -\bar{s}(x)]
\end{eqnarray}
and for the $\nu(\bar{\nu})$-neutron scattering $F_{2,3}(x)$ are given by
\begin{eqnarray}\label{eq:pdf2}
F_{2n}^{\nu }(x) & = & 2 x [u(x) + s(x) + \bar{d}(x) +\bar{c}(x)]\;;\;\;F_{2n}^{\bar\nu}(x)  =  2 x [d (x)+ c(x) + \bar{u}(x) +\bar{s}(x)]\nonumber\\
x F_{3n}^{\nu}(x) & = & 2 x [u(x) + s (x)- \bar{d}(x) -\bar{c}(x)]\;;\;\;
x F_{3n}^{\bar\nu}(x)  =  2 x [d(x) + c(x) - \bar{u}(x) -\bar{s}(x)]. 
\end{eqnarray}
For an isoscalar nucleon target, we use
\begin{equation}
 F_{iN}=\frac{F_{ip}+F_{in}}{2},~~~i=1-5
\end{equation}
The remaining two structure functions $F_{1N}(x)$ and $F_{5N}(x)$ at the leading order are written using Callan-Gross~\cite{Callan:1969uq} and Albright-Jarlskog~\cite{Albright:1974ts} relations, respectively as:
%
\begin{eqnarray}
 F_{1}(x)&=&\frac{F_2(x)}{2 x}\;;\;\;
 F_{5}(x)=\frac{F_2(x)}{2 x} \nonumber
\end{eqnarray}
The parton distribution functions (defined in Eqs.\ref{parton_wk}, \ref{eq:pdf1} and \ref{eq:pdf2}) for the nucleon have been determined by various groups and they are known in the literature by the acronyms MRST~\cite{Martin:1998sq}, 
 GRV~\cite{Gluck:1998xa}, 
 GJR~\cite{Gluck:2007ck}, MSTW~\cite{Martin:2009iq}, ABMP~\cite{Alekhin:2016uxn}, ZEUS~\cite{Chekanov:2002pv}, HERAPDF~\cite{Zhang:2015tuh}, 
 NNPDF~\cite{DelDebbio:2007ee}, CTEQ~\cite{Nadolsky:2008zw}, CTEQ-Jefferson Lab (CJ)~\cite{Accardi:2009br}, MMHT~\cite{Harland-Lang:2014zoa}, etc. In the present work the numerical results are presented using MMHT~\cite{Harland-Lang:2014zoa} nucleon parton distribution functions.

In the present formalism we shall treat up, down and strange quarks to be massless and the charm quark to be a massive object. For the case of charm quark density distribution at the leading order, we use the Cabibbo-Kobayashi-Maskawa (CKM) mixing matrix which is given by
$$\begin{bmatrix}
 d^\prime \\
 s^\prime\\
 b^\prime
\end{bmatrix}=\begin{bmatrix}V_{ud}&V_{us}&V_{ub}\\
V_{cd}&V_{cs}&V_{cb}\\
 V_{td}&V_{ts}&V_{tb}\end{bmatrix}\;\begin{bmatrix} d\\s\\b \end{bmatrix}
$$
where $\theta_C$ is the Cabibbo angle 
and $|V_{ij}|^2$ with $i$ and $j$ as the flavor of quarks represents the probability that a quark of flavor $j$ will decay into a quark of flavor $i$. Hence, the probability density for charm quark (dropping the suppressed bottom quark initiated contributions) is given by~\cite{Kretzer:2002fr, Kretzer:2001tc}:
\begin{equation}
 s'=s\;|V_{cs}|^2+d\;|V_{cd}|^2=s\;\cos^2\theta_C+d\;\sin^2\theta_C.
\end{equation}

Till now we have discussed the structure functions in the leading order. However, in QCD,  partons present inside the nucleon may interact among themselves via gluon exchange. The incorporation of contribution from gluon emission induces the $Q^2$ dependence 
of the nucleon structure functions, i.e. Bjorken scaling is violated. The $Q^2$ evolution of
structure functions is determined by the DGLAP evolution equation~\cite{Altarelli:1977zs}. In the next section, we will discuss 
the higher order perturbative QCD corrections. 

\section{QCD Corrections}\label{qcd}
\subsection{NLO Evolution}
\label{nlo}
In the naive parton model in the limit of $Q^2\to \infty,~\nu\to \infty$ with $x\to\textrm{``a finite value''}$, nucleon structure functions would only be the function of Bjorken variable ($x$). The probability of the gluon emission due to the interaction of nucleons is related to the strong coupling constant $\alpha_s(Q^2)$, which changes with the value of $Q^2$. For example, in the limit of $Q^2\to\infty$, the strong coupling constant $\alpha_s(Q^2)$ becomes very small and, therefore, the higher order terms can be neglected in comparison to the leading order(LO) terms. While for a finite value of $Q^2$, $\alpha_s(Q^2)$ is large and higher order terms such as next-to-leading order (NLO) gives a significant contribution. The $Q^2$ evolution of structure functions is determined by the DGLAP evolution equation~\cite{Altarelli:1977zs}.
Hence, one may express the nucleon structure functions in terms of the convolution of coefficient function ($H_f\;;\;(f=q,g)$) with the density distribution of partons ($f$) inside the nucleon as
\begin{equation}\label{f2_conv}
 x^{-1} F_{i} (x) = \sum_{f=q,g} H_{f}^{(n)}(x) \otimes f(x)\; ,
\end{equation}
where $i=1-5$, superscript $n=0,1,2,...$ for N$^{(n)}$LO evolution and symbol $\otimes$ is the Mellin convolution. 
To obtain the convolution of coefficient functions with
parton density distribution, we use the following expression
\begin{equation}
 H_f(x)\otimes f(x) =\int_x^1 H_f(y)\; f\left(\frac{x}{y} \right) {dy \over y}
\end{equation}
This Mellin convolution turns into simple multiplication in the N-space.
The parton coefficient function are generally expressed as 
\begin{eqnarray}
 H_{f}(x,Q^2)=\underbrace{H_{f}^{(0)}}_{LO}+\underbrace{\frac{\alpha_s(Q^2)}{2\pi}H_{f}^{(1)}}_{NLO}+\underbrace{\left(\frac{\alpha_s(Q^2)}{2\pi}\right)^2\;H_{f}^{(2)}}_{NNLO}+...
\end{eqnarray}
For example, the dimensionless structure functions in the case of massive charm are given by~\cite{Kretzer:2002fr, Kretzer:2003iu}
\begin{eqnarray}\label{eq_chm}
 F_i^c(x,Q^2)&=&(1-\delta_{i4})\cdot s^\prime(\bar{\xi}, \mu^2)+\frac{\alpha_s(\mu^2)}{2\pi}\Big\{\int_{\bar{\xi}}^1 \frac{dy^\prime}{y^\prime}
 \Big[H_i^q\Big( y^\prime, \frac{Q^2}{\mu^2}, \lambda \Big)s^\prime\Big(\frac{\bar{\xi}}{y^\prime}, \mu^2\Big)  \nonumber\\
 &&+ H_i^g\Big( y^\prime, \frac{Q^2}{\mu^2}, \lambda \Big)g^\prime\Big(\frac{\bar{\xi}}{y^\prime}, \mu^2\Big)   \Big]\Big\},
\end{eqnarray}
where at the leading order
\begin{equation}
 H_{f}^{(0)}=(1-\delta_{i4}),
\end{equation}
and the terms at the next-to-leading order ($H_i^q$ and $H_{i}^g$ ; ($i=1-5$)) with strong coupling constant $\frac{\alpha_s(\mu^2)}{2\pi}$ gives finite contribution. From the above expression, it may be noticed that though at the leading order $F_{4}(x)=0$ but when we have taken NLO terms into account, we
obtain a nonzero contribution for $F_{4}(x)$ i.e. $F_{4}(x)\ne 0$. 
In Eq.\ref{eq_chm}, $g^\prime$ is the gluonic density which is given by~\cite{Kretzer:2002fr, Kretzer:2001tc}:
\begin{eqnarray}
 g^\prime&=&g\cdot \cos^2\theta_C + g\cdot \sin^2\theta_C,
\end{eqnarray}
$H_i^q$ and $H_i^g$; ($i=1-5$) are respectively the fermionic and gluonic coefficient functions at NLO which are taken following Ref.~\cite{Kretzer:2003iu}. In Eq.\ref{eq_chm}, $\bar\xi$ is the slow rescaling variable (discussed in the subsection~\ref{TMC}) and the variables $\lambda$ and $y^\prime$ are defined as
\begin{equation}
 \lambda=\frac{Q^2}{(Q^2+m_c^2)}\;,\;\;y^\prime=\frac{\bar\xi}{y}
\end{equation}
where $m_c$ is the charm quark mass.
Notice that in the case of massless quarks, the variable $\lambda$ will be equal to 1.

\subsection{Target Mass Corrections (TMC) Effect}
\label{TMC}
 The target mass correction is a nonperturbative effect, which comes into the picture at lower $Q^2$. 
 At finite values of $Q^2$, the mass of the target nucleon and the quark masses modify the Bjorken variable $x$ with the light cone momentum fraction. For the massless quarks, the parton light cone momentum fraction is given by the Nachtmann variable $\xi$ which is related to the Bjorken variable $x$ as
 \begin{equation}\label{nach}
 \xi=\frac{2x}{1+\rho}\;;\;\;\rho=\sqrt{1+4 \mu x^2} \;,\;\mu=\frac{M_N^2}{Q^2}.
 \end{equation}
 The Nachtmann variable $\xi$ depends only on the hadronic mass and will not have corrections due to the masses of final state quarks. However, for the massive partons, the Nachtmann variable $\xi$ gets modified to the slow rescaling variable $\bar\xi$. These variables $\xi$ and $\bar\xi$ are related as:
\begin{equation}\label{snach}
    \bar\xi=\xi \left( 1+\frac{m_c^2}{Q^2}\right)=\frac{\xi}{\lambda}
\end{equation}
 It may be noticed from Eqs.~\ref{nach} and \ref{snach} that the Nachtmann variable corrects the Bjorken variable for the effects of hadronic mass while the generalized variable $\bar\xi$ further corrects $\xi$ for the effects of the partonic masses.
The simplified expression of target mass corrected structure functions for massless quarks ($u,~d,$ and $s$ in our case) are given by
\begin{eqnarray}
\label{eqn:tmc}
F_{1N}^{TMC}(x,Q^2)&=& \frac{x}{\xi \rho}\;F_{1N}^{0}(\xi,Q^2)+\frac{\mu x^2}{\rho^2} \;h_2(\xi,Q^2)+
\frac{2 \mu^2 x^3}{\rho^3}\;g_2(\xi,Q^2)\nonumber\\
 F_{2N}^{TMC}(x,Q^2)&=&\frac{x^2}{\rho^3 \xi^2}\; F_{2N}^{0}(\xi,Q^2)+\frac{6 \mu x^3}{\rho^4}\;h_2(\xi,Q^2)+\frac{12 \mu^2 x^4}{\rho^5}\;g_2(\xi,Q^2)\nonumber\\
 F_{3N}^{TMC}(x,Q^2)&=&\frac{x}{\rho^2 \xi}\;F_{3N}^{0}(\xi,Q^2)+\frac{2 \mu x^2}{\rho^3}\;h_3(\xi,Q^2)\nonumber\\
 F_{4N}^{TMC}(x,Q^2)&=& \frac{\mu^2 x^3}{\rho^3} \;F_{2N}^{0}(\xi,Q^2)+\frac{1}{1+\mu \xi^2}\;F_{4N}^{0}(\xi,Q^2)-\frac{2 \mu x^2}{\rho^2}\;F_{5N}^{0}(\xi,Q^2)\nonumber\\
&-&\frac{2\mu^2 x^4 (2-\mu\xi^2)}{\rho^4}\;h_2(\xi,Q^2)+\frac{\mu x^2}{\rho^3}\;h_5(\xi,Q^2)+\frac{2 \mu^2 x^3 (1-2\mu x^2)}{\rho^5}\;g_2(\xi,Q^2)\nonumber\\
 F_{5N}^{TMC}(x,Q^2)&=& -\frac{\mu x^2}{\rho^3 \xi}\;F_{2N}^{0}(\xi,Q^2)+\frac{x}{\rho^2 \xi}\;F_{5N}^{0}(\xi,Q^2)+\frac{2 \mu x^2 (1-\mu \xi x)}{\rho^4} \; h_2(\xi,Q^2)\nonumber\\
 &+&\frac{\mu x^2}{\rho^3}\; h_5(\xi,Q^2)+\frac{6 \mu^2 x^3}{\rho^5}\;g_2(\xi,Q^2)\nonumber\\
\end{eqnarray}
While for the case of massive charm the following expressions are used:
\begin{eqnarray}\label{eqn:tmcc}
 F_{1N,c}^{TMC}(x,Q^2)&=&\frac{x}{\xi \rho}\;F_{1N,c}^{0}(\xi,Q^2)+\frac{2\mu x^2}{\lambda \rho^2}\;H_2(\xi,Q^2)+\frac{4\mu^2 x^3}{\rho^3}\;G_2(\xi,Q^2)\nonumber\\
  F_{2N,c}^{TMC}(x,Q^2)&=&\frac{2 x^2}{\lambda \rho^3 \xi}\;F_{2N,c}^{0}(\xi,Q^2)+\frac{12\mu x^3}{\lambda \rho^4}\;H_2(\xi,Q^2)+\frac{24 \mu^2 x^4}{\lambda \rho^5}G_2(\xi,Q^2)\nonumber\\
 F_{3N,c}^{TMC}(x,Q^2)&=& \frac{2x}{\rho^2 \xi}\;F_{3N,c}^{0}(\xi,Q^2)+\frac{4 \mu x^2}{\rho^3}\;H_3(\xi,Q^2)\nonumber\\
  F_{4N,c}^{TMC}(x,Q^2)&=& \frac{2 \mu^2 \xi^2 x^2}{\lambda \rho^2(1+\mu \xi^2)}\;F_{2N,c}^{0}(\xi,Q^2)+\frac{1}{\mu \xi^2}\;F_{4N,c}^{0}(\xi,Q^2)-\frac{2\mu x \xi}{\rho(1+\mu \xi^2)}\;F_{5N,c}^{0}(\xi,Q^2)\nonumber\\
  &-&\frac{4 \mu^2 x^4 (2-\mu \xi^2)}{\lambda \rho^4}\;H_2(\xi,Q^2)+\frac{2\mu x^2}{\rho^3}\;H_5(\xi,Q^2)+\frac{4 \mu^2 x^3(1-2\mu x^2)}{\lambda \rho^5}\;G_2(\xi,Q^2)\nonumber\\
 F_{5N,c}^{TMC}(x,Q^2)&=& \frac{-2\mu x^2}{\lambda \rho^3}\;F_{2N,c}^{0}(\xi,Q^2)+\frac{x}{\rho^2 \xi}\;F_{5N,c}^{0}(\xi,Q^2)+\frac{4\mu x^2 (1-\mu x \xi)}{\lambda \rho^4}\; H_2(\xi,Q^2)\nonumber\\
 &+&\frac{2 \mu x^2}{\rho^3}\;H_5(\xi,Q^2)+\frac{12 \mu^2 x^3}{\rho^5}\;G_2(\xi,Q^2)\nonumber\\
\end{eqnarray}
In the expressions of TMC corrected structure functions (Eqs.\ref{eqn:tmc} and \ref{eqn:tmcc}), $F_{iN}^{0}$ and $F_{iN,c}^{0};~~(i=1-5)$ represent the bare structure functions, i.e., without the TMC effect.
One may also notice from Eqs.\ref{eqn:tmc} and \ref{eqn:tmcc} that even at the leading order, there will be nonzero contribution from the target mass corrected 
structure function $F_{4N}^{TMC}(x,Q^2)$ in the case of massless as well as massive quarks.
In the above expressions following integrations have been used~\cite{Kretzer:2003iu}:
\begin{eqnarray}
h_i(\xi,Q^2) &=& \int_\xi^1 d y \frac{F_{i}^0(y,Q^2)}{y}*\phi_i\;;\;\;
g_2(\xi,Q^2)=\int_{\bar\xi}^1\;dy \; h_2(y,Q^2)\nonumber\\
 H_i(\xi,Q^2)&=&\int_{\bar\xi}^1\;dy \frac{F_{iN,c}(y,Q^2)}{y}\;;\;\;
 G_2(\xi,Q^2)=\int_{\bar\xi}^1\;dy H_i(y,Q^2) \nonumber
\end{eqnarray}
The variable $\phi_i$ is given in Table~\ref{phi_tb}.
\begin{table}
 \begin{tabular}{|c|c|c|c|c|c|}\hline
 Variable &i=1&i=2&i=3&i=4&i=5\\\hline
$\phi_i$  &2&$\frac{1}{y}$&1&4&2\\\hline
 \end{tabular}
 \caption{Constant variable $\phi_i$.}
 \label{phi_tb}
\end{table}

\subsection{Dynamical Higher Twist(HT: twist-4) Effect:}
\label{HT}
For lower values of $Q^2$, a few GeV$^2$ or less, nonperturbative 
phenomena becomes important for a precise modeling of cross sections. In the present work, besides TMC effect we have also taken into account the dynamical higher twist (HT) effect which deals with the interaction of struck quark with other surrounding quarks via gluon exchange. HT corrections are suppressed by the power of $\left({1 \over Q^2} \right)^n$, where $n=1,2,....$ and therefore,  pronounced in the region of low $Q^2$ and high $x$. For high $Q^2$ and low $x$ it becomes negligible like the TMC effect. In the formalism of the operator product expansion (OPE)~\cite{Wilson:1969zs, Shuryak:1981kj}, unpolarized structure 
functions can be expressed in terms of powers of $1/Q^2$ (power corrections):

\begin{equation}
F_{i}(x,Q^2) = F_{i}^{j = 2}(x,Q^2)
+ {{\cal H}_{i}^{j = 4}(x) \over Q^2}  + .....   \;\;\; i=1,2,3, 
\label{eqn:ht}
\end{equation}
where the first term ($j=2$) is known as the twist-two or leading twist (LT) term, and it corresponds to the scattering off a free quark. 
 This term obeys the Altarelli-Parisi equations and is expressed in terms of PDFs. It is responsible for the evolution 
of structure functions via perturbative QCD $\alpha_s(Q^2)$ corrections. 
The term corresponding to $j=4$ is known as the twist-4 or higher twist term and it reflects the multiparton correlations. We have observed in our earlier study~\cite{Zaidi:2019asc} that the scattering cross section obtained with TMC and HT corrections at NLO have negligible difference from the results obtained at NNLO with the TMC effect only.
 \section{Results and Discussion}\label{results}
\begin{figure}
 \includegraphics[height=16 cm, width=0.95\textwidth]{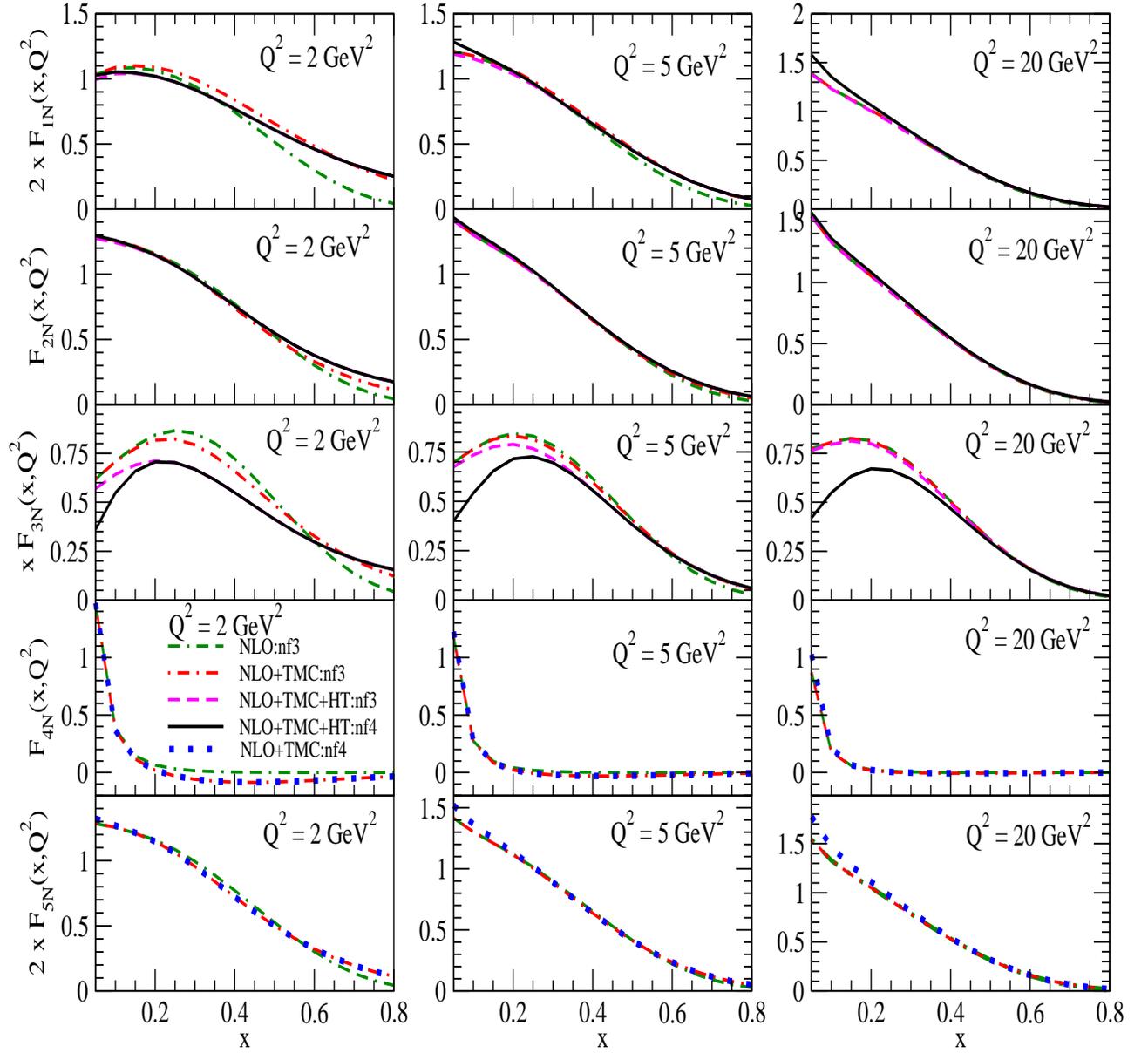}
 \caption{ Results for the free nucleon structure functions $F_{iN}(x,Q^2)$;$(i=1-5)$(Top to Bottom)
  at the different values of $Q^2$ viz. 2, 5 and 
 10 GeV$^2$(Left to Right) are shown. These results are obtained at NLO by using MMHT nucleon PDFs
 parameterization~\cite{Harland-Lang:2014zoa}. The results are shown without the TMC effect (double dashed-dotted line), 
 with the TMC effect in the 3-flavor(nf3) scheme (dashed-dotted line) as well as four flavor(nf4) scheme(dotted line), with TMC and HT effects in the 3-flavor(nf3) scheme (dashed line) as well as four flavor(nf4) scheme(solid line).}
 \label{fig:sf_free}
\end{figure}

\begin{figure}
 \includegraphics[height=8 cm, width=0.9\textwidth]{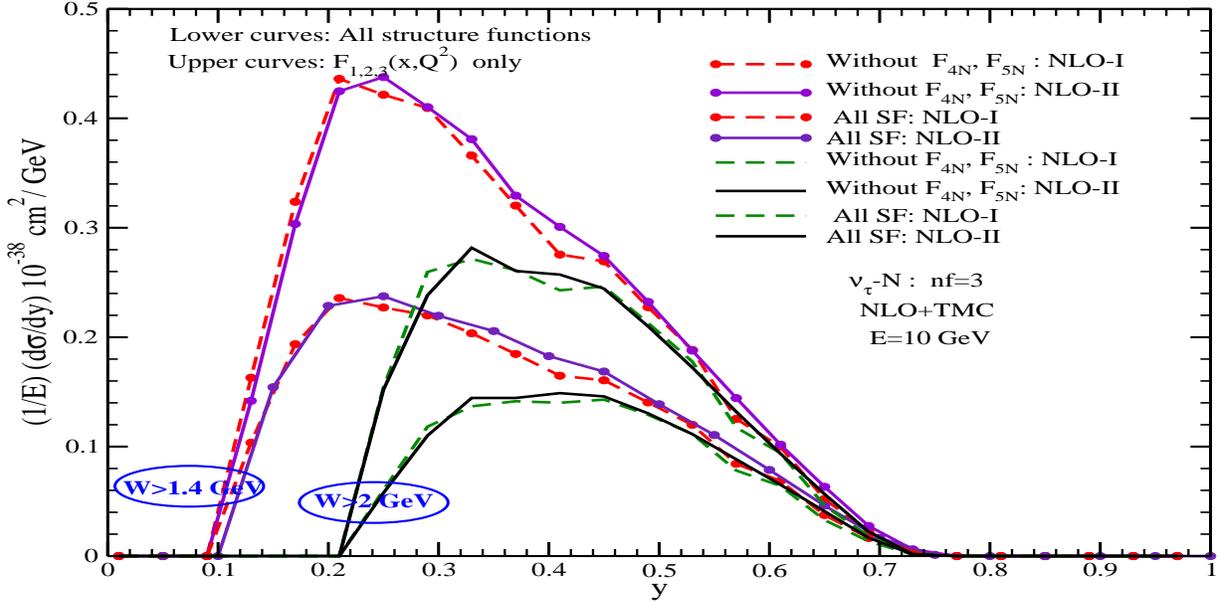}
 \caption{$\frac{1}{E}\frac{d\sigma}{dy}$ vs $y$ at $E=10$ GeV for the $\nu_\tau - N$ DIS. These results are obtained at NLO following the approach of NLO evolution by {\bf (I)}  Vermaseren et al. and Moch et al.~\cite{Vermaseren:2005qc, Moch:2004xu, Moch:2008fj} (labeled as NLO-I) and {\bf (II)} Kretzer et al.~~\cite{Kretzer:2002fr} (labeled as NLO-II).
   These results are presented with a CoM($W$) cut of 1.4 GeV (LHS) and 2 GeV (RHS). The effect of the target mass correction~\cite{Kretzer:2003iu} is included. The upper curves represent the cross section when we have taken into account only the three structure functions $F_{iN}(x,Q^2);~~i=1-3$ while the lower curves represent the case when we have also included $F_{iN}(x,Q^2);~~i=4-5$ i.e. all the five structure functions in the numerical calculations.}
 \label{fig:d2sigf}
\end{figure}

\begin{figure}
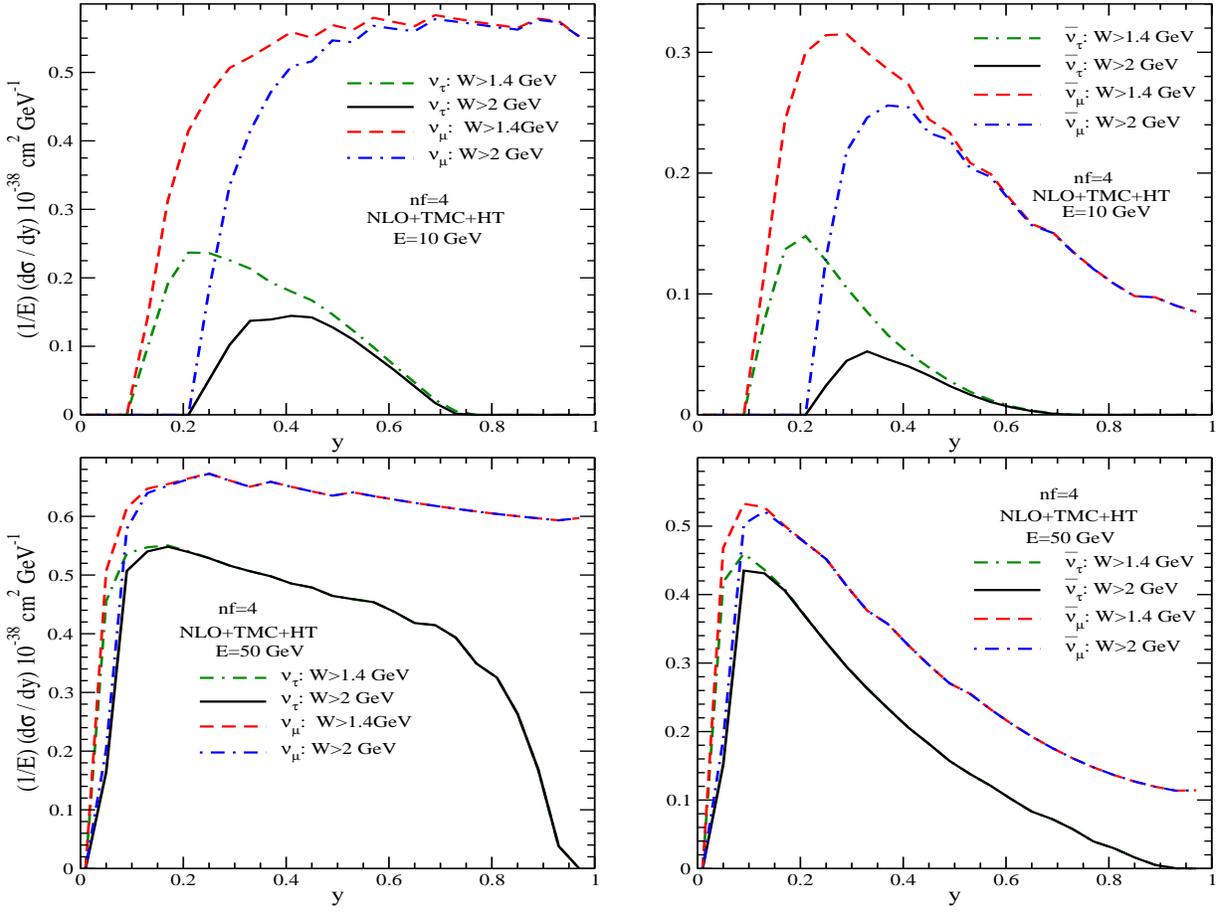

 \includegraphics[height=6 cm, width=0.9\textwidth]{dsigma_over_dy_10gev_v1.eps}\\
  \includegraphics[height=6 cm, width=0.9\textwidth]{dsigma_over_dy_50gev_v1.eps}
  \caption{$\frac{1}{E}\frac{d\sigma}{dy}$ vs $y$ at $E$=10 GeV and 50 GeV are shown for the muon and tauon type 
  neutrinos and antineutrinos. These results are obtained in four flavor scheme with $W>1.4$ GeV and $W>2$ GeV. 
   The effects of TMC~\cite{Kretzer:2003iu} and HT~\cite{Dasgupta:1996hh} are included.
  The $\nu_\mu$(Left panels) and ${\bar\nu}_\mu$(Right panels) results are shown with dashed($W > 1.4$ GeV) and 
  dash-dotted($W > 2$ GeV) lines, while the $\nu_\tau$(Left panels) and ${\bar\nu}_\tau$(Right panels) results are shown with double dash-dotted ($W > 1.4$ GeV)  and 
  solid ($W > 2$ GeV) lines.
  }
  \label{fig:dsigdy1050}
  \end{figure}

\begin{figure}
 \includegraphics[height=8 cm, width=0.9\textwidth]{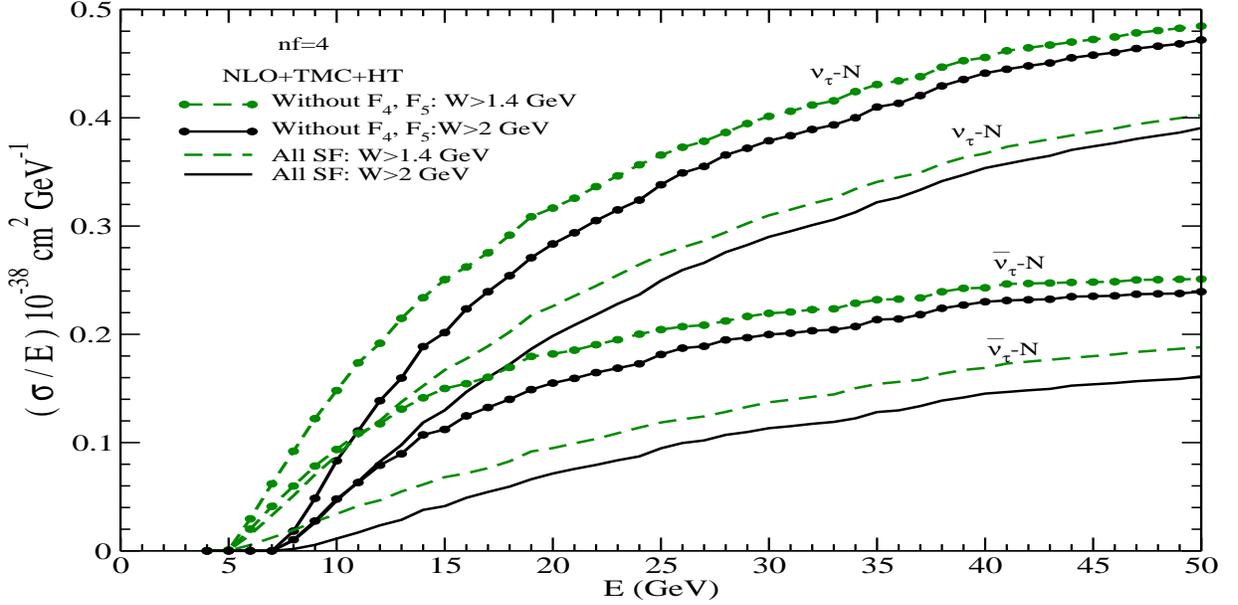}
 \caption{Results for the total scattering cross section $\frac{\sigma}{E}$ vs $E$ with center of mass energy ($W$) cut of 1.4 GeV and 2 GeV for $\nu_\tau-N$ and $\bar\nu_\tau-N$ DIS are shown. These results are obtained at NLO by using MMHT nucleon PDFs parameterization~\cite{Harland-Lang:2014zoa} with TMC effect~\cite{Kretzer:2003iu} and HT corrections~\cite{Dasgupta:1996hh} in the four flavor scheme.}
 \label{fig:sigma_f4f5}
\end{figure}

\begin{figure}
 \includegraphics[height=16 cm, width=0.9\textwidth]{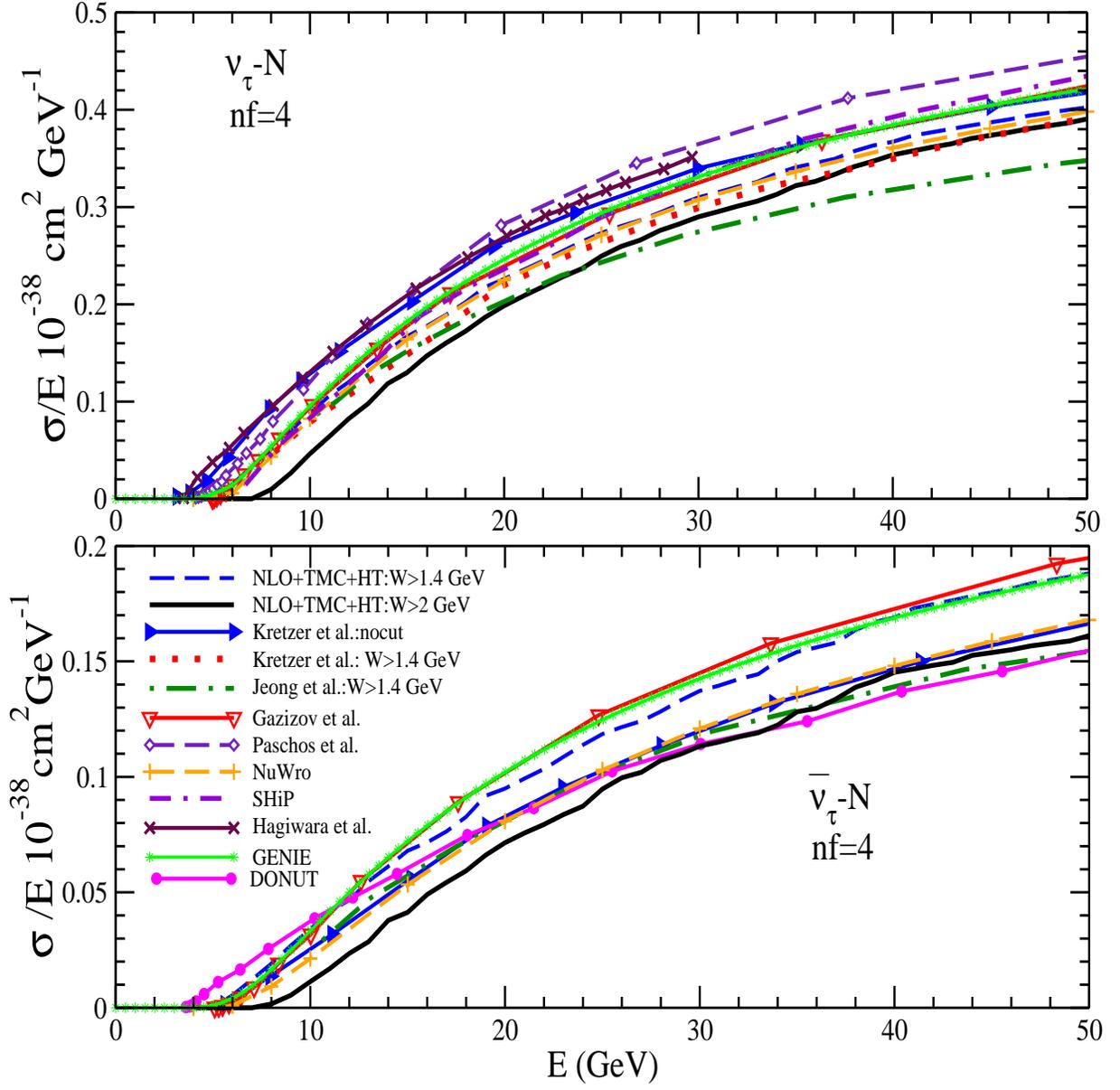}
 \caption{$\frac{\sigma}{E}$ vs $E$ with a center of mass energy ($W$) cut of 1.4 GeV(dashed line) and 2 GeV(solid line), for tau type neutrinos(Top panel)
 and antineutrinos(Bottom panel) with the TMC~\cite{Kretzer:2003iu} and higher twist~\cite{Dasgupta:1996hh} effects. These results are  compared with the results of different models available in the literature~\cite{Kretzer:2002fr, Jeong:2010nt, Hagiwara:2003di, Conrad:2010mh, Paschos:2001np, Gazizov:2016, Li:2017dbe, Anelli:2015pba} as well as with the Monte Carlo generators GENIE~\cite{GENIE} and NuWro\cite{NuWro}.}
 \label{fig:sig_comp}
\end{figure}

\begin{figure}
 \includegraphics[height=8 cm, width=0.9\textwidth]{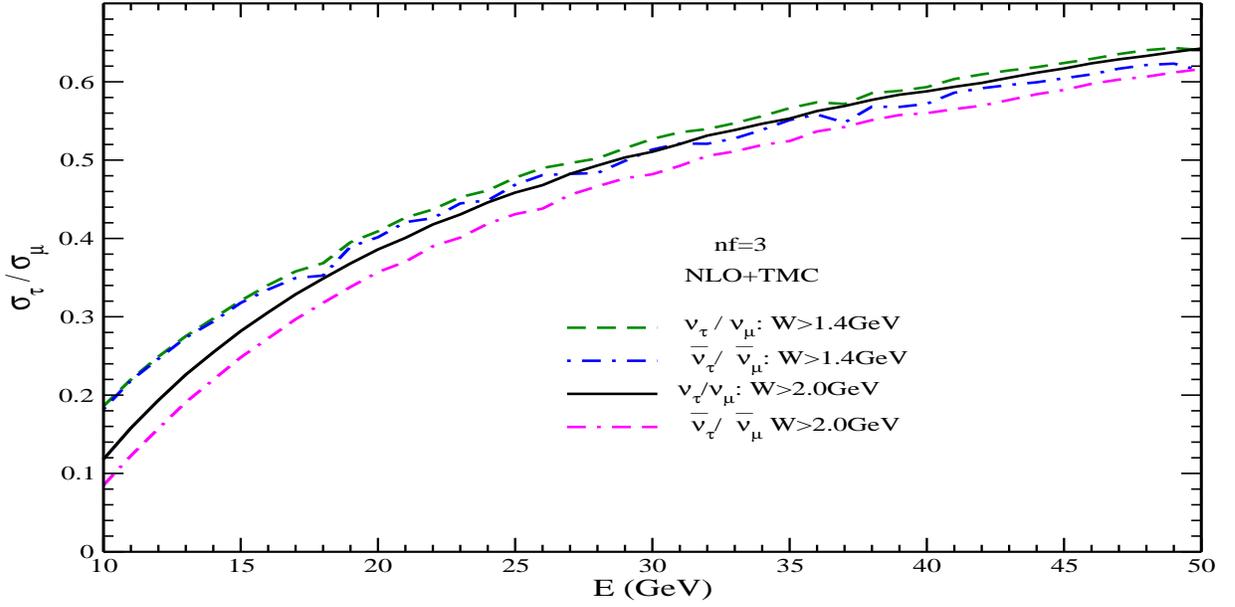}
 \caption{Ratio of the total scattering cross section $\frac{\sigma_{\nu_\tau}}{\sigma_{\nu_\mu}}$ vs $E$ are shown with $W > 1.4$ GeV and $W > 2$ GeV for $\nu_\tau - N$ and ${\bar\nu}_\tau - N$ DIS. Dashed and dashed-dotted lines represent the results with $W>1.4$ GeV while the solid and double dashed-dotted lines represent the results with $W>2$ GeV for neutrinos and antineutrinos, respectively.  The effect of TMC~\cite{Kretzer:2003iu} is also included.}
 \label{fig:sig_ratio}
\end{figure}

 In this section, we have presented the results for the free nucleon structure functions, the differential and the total scattering cross sections obtained by using the formalism discussed in the previous section. Numerical results for the various structure functions and the differential and total scattering cross sections are presented in Figs.~\ref{fig:sf_free}-\ref{fig:sig_ratio}. All the results are presented at NLO incorporating the following considerations:
 \begin{itemize}
 \item The target mass correction effect is calculated for the massive as well as massless quarks using Eqs.~\ref{eqn:tmc} and \ref{eqn:tmcc} in the evaluation of nucleon structure functions $F_{iN}(x,Q^2);~~i=1-5$ and the differential and total scattering cross sections.
  \item The higher twist (HT) effect is calculated in terms of function ${\cal H}_{i}^{j = 4}(x)$ using Eq.~\ref{eqn:ht} in the  evaluation of $F_{iN}(x,Q^2);~~i=1-3$. 
  \item The results are presented in the three(nf3)- and four(nf4)- flavor schemes. The effect of massive charm quark has been taken into account.
  \item All the results are presented using MMHT PDFs parameterization of Harland-Lang et al.~\cite{Harland-Lang:2014zoa}.
  \item A cut in $Q^2$ of $Q^2 \ge 1$~GeV$^2$ has been used to calculate the cross sections using DIS formalism.
\item The lepton mass effect has been shown explicitly by comparing $\nu_\mu$ vs $\nu_\tau$ induced differential and the total scattering cross sections.
\item The effect of CoM energy($W$) cuts on the nucleon structure functions, as well as on the differential and the total scattering cross sections has been studied.
\item The uncertainty in the nucleon structure functions and scattering cross section arising due to different approaches of NLO evolution has also been studied. 
 \end{itemize}

 In Fig.~\ref{fig:sf_free}, the results for the free nucleon structure functions $2xF_{1N}(x,Q^2)$, $F_{2N}(x,Q^2)$, $xF_{3N}(x,Q^2)$, $F_{4N}(x,Q^2)$ and $2xF_{5N}(x,Q^2)$ 
  (from the top to bottom) are shown at the different values of $Q^2$.
   The results shown in the left panel correspond to $Q^2=2$ GeV$^2$, in the middle panel correspond to $Q^2=5$ GeV$^2$ and in the right panel corresponds to $Q^2=20$ GeV$^2$. The nucleon structure functions are presented at NLO 
   without the TMC effect (double dash-dotted line), with the TMC effect in 3-flavor(dash-dotted line:nf3) and 4-flavor(dotted line:nf4) schemes, with TMC and HT effects in 3-flavor(dashed line: nf3) and 4-flavor(solid line:nf4) schemes. From the figure, it may be noticed that the TMC effect is dominant in the region of high $x$ and low $Q^2$ and it becomes small at low $x$ and high $Q^2$. Quantitatively, the TMC effect is found to be different in $F_{2N}(x,Q^2)$ from $F_{1N}(x,Q^2)$ while 
  the TMC effect in $F_{5N}(x,Q^2)$ is similar to the effect in $F_{2N}(x,Q^2)$. However, in the case of $F_{4N}(x,Q^2)$ the whole contribution arises in the leading order due to the TMC effect at mid and high $x$. $F_{4N}(x,Q^2)$ contributes to the cross section due to large lepton mass, and  contributes only in the region of $x \le 0.2$. We find that at NLO, $F_{4N}(x,Q^2)$ becomes almost negligible in the region of $x>0.2$ when TMC effect is not incorporated but with the inclusion of TMC effect a nonzero though small contribution in the region of high $x$ and low $Q^2$ has been found. The difference in the results of free nucleon structure functions $F_{iN} (x, Q^2);~~ (i = 1-5)$ evaluated at NLO with and without the TMC effect at $x=0.3$ is $5\%(3\%)$ in $F_{1N} (x, Q^2)$, $2\%(<1\%)$ in $F_{2N} (x, Q^2)$, $7\%(\sim 3\%)$ in $F_{3N} (x, Q^2)$ and $4\%(\sim 2\%)$ in $F_{5N} (x, Q^2)$ for $Q^2=2(5)$ GeV$^2$.
  
   In the case of first three structure functions ($F_{iN}(x,Q^2);~~i=1-3$), the HT effect has also been included. This is found to be comparatively small in $F_{1N}(x,Q^2)$ and $F_{2N}(x,Q^2)$ than in $F_{3N}(x,Q^2)$. Due to higher twist corrections, we have observed a decrease in the value of  $F_{3N}(x,Q^2)$, which becomes small with the increase in $Q^2$. 
   To show the effect of massive charm on the nucleon structure functions, we have compared the results obtained with the TMC and HT effects for the three flavor of massless quarks (nf3) with the results when additional contribution from massive charm quark (nf4) have also been considered. It is found that
   massive charm effect is almost negligible in the kinematic region of low $Q^2$ and high $x$ while it increases with the increase in $Q^2$ and is appreciable at low $x$. With the HT effect included in the evaluation of $F_{iN} (x, Q^2);~~ (i = 1-3)$, we find at $Q^2=2(5)$ GeV$^2$ and $x=0.3$ there is an effect of $2\%(3\%)$ in $F_{1N} (x, Q^2)$, $<1\%(<1\%)$ in $F_{2N} (x, Q^2)$ and $\sim 20\%(7\%)$ in $F_{3N} (x, Q^2)$. With the increase in $x(=0.8)$ the effect of HT further increases to $27\%(6\%)$ in $F_{1N} (x, Q^2)$, $35\%(\sim 20\%)$ in $F_{2N} (x, Q^2)$ and $\sim 21\%(11\%)$ in $F_{3N} (x, Q^2)$.  We find that the effect due to HT is somewhat larger for $F_{3N}(x,Q^2)$ at low $x$ and low $Q^2$ which becomes small with the increase in $Q^2$.  
 
 In Fig.\ref{fig:d2sigf}, the results for the differential cross section $\frac{1}{E}\frac{d\sigma}{dy}$ vs $y$ at $E=10$ GeV are shown for $\nu_\tau-N$ DIS process. These results are obtained by incorporating the TMC effect with different cuts on the CoM energy viz. $W>1.4$ GeV and $W>2$ GeV.
 The upper curves represent the cross section when we have taken into account only the three structure functions $F_{iN}(x,Q^2);~~i=1-3$ while the lower curves represent the case when we have included the contributions from all the five structure functions in the numerical calculations. From the figure, it may be observed that the inclusion of $F_{4N}(x,Q^2)$ and $F_{5N}(x,Q^2)$ results in a suppression of the scattering cross section. We find that the suppression is predominantly due to the inclusion of $F_{5N}(x,Q^2)$(not shown here explicitly).
  The effect of CoM cut on the differential cross section has also been presented. It is found that a cut on the CoM at higher W  removes the lower region of inelasticity ($y$) as well as reduces the strength of the differential scattering cross section. In this figure, we have also made a 
comparison of the two different NLO approaches given by: {\bf (I)} Vermaseren et al.~\cite{Vermaseren:2005qc} and Moch et al.~\cite{Moch:2004xu, Moch:2008fj} (dashed lines without and with solid circles)
and {\bf (II)} Kretzer et al.~\cite{Kretzer:2002fr} (solid lines without and with solid circles). One may notice a good agreement between the results obtained by using these two approaches. 
 Therefore, it may be concluded that the choice of the different approaches for the evolution of nucleon structure functions at NLO would not make much difference in the differential scattering cross sections in the case of free nucleon.
  
  In Fig.~\ref{fig:dsigdy1050}, the results are presented for the $\nu_l-N$ and $\bar\nu_l-N$; ($l=\mu,\tau)$ differential scattering cross sections ($\frac{1}{E}\frac{d\sigma}{dy}$) at $E=10$ GeV and 50 GeV. 
  These results are obtained by taking into account TMC~\cite{Kretzer:2003iu} and HT~\cite{Dasgupta:1996hh} effects 
   in the four flavor scheme at NLO. These results depict the kinematic effect in order to understand the lepton mass effect(m$_\mu$ vs m$_\tau$) and the effect of CoM energy cut on the scattering cross section. Due to the threshold effect as well as the appearance of 
   additional structure functions $F_{4N}(x,Q^2)$ and $F_{5N}(x,Q^2)$, the scattering cross section for $\nu_\tau$ and $\bar\nu_\tau$ induced processes are smaller in  magnitude than in the case of 
  $\nu_\mu$ and $\bar\nu_\mu$ induced processes. Quantitatively, at $E=10$ GeV there is a suppression of $43\%(51\%)$ at $y=0.2$, and $74\%(88\%)$ at $y=0.5$ in the cross section for $\nu_\tau-N(\bar\nu_\tau-N)$ DIS process from the case of $\nu_\mu-N(\bar\nu_\mu-N)$ when a cut of $W>1.4$ GeV is incorporated. It implies that in the region of high $y$, the contribution from the $\nu_\tau/\bar\nu_\tau$ events to the scattering cross section becomes almost negligible.
  The effect of CoM energy cut is significantly visible in the region of low $y$ at $E=10$ GeV, however, with the increase in energy, i.e. at $E=50$ GeV it becomes small. For example, at $E=10$ GeV the difference between the results with $W>1.4$ GeV and $W>2$ GeV is $34\%(30\%)$ for $\nu_\mu(\bar\nu_\mu)$ and $55\%(58\%)$ for $\nu_\tau(\bar\nu_\tau)$ at $y=0.3$, however, this difference becomes negligible at $E=50$ GeV.
  
   In Fig.~\ref{fig:sigma_f4f5}, the results for the total scattering cross section $\sigma$ vs neutrino energy ($E$) are presented with $W>1.4$ GeV and $W>2$ GeV. These results are evaluated by taking into account TMC~\cite{Kretzer:2003iu} and HT~\cite{Dasgupta:1996hh} effects in the four flavor scheme in the presence of $F_{4N}(x,Q^2)$ and $F_{5N}(x,Q^2)$ as well as when these two structure functions are switched off. The results shown by the dashed($W>1.4$ GeV) and solid($W>2$ GeV) lines represent the case when all structure functions are included, while the dashed line($W>1.4$ GeV) with dots and solid lines with dots($W>2$ GeV) represent 
  the case when $F_{4N}(x,Q^2)$ and $F_{5N}(x,Q^2)$ are switched off. From the figure, it may be observed that the cross section for  the antineutrino scattering process is approximately $61\%$ at E=10 GeV, $58\%$ at E=20 GeV and $53\%$ at E=50 GeV of the neutrino induced scattering process when all the five structure functions are considered and a CoM energy cut of 1.4 GeV is applied. Here one may also notice that the inclusion of $F_{4N}(x,Q^2)$ and $F_{5N}(x,Q^2)$ leads to a suppression in the scattering cross section from the cross sections calculated without $F_{4N}(x,Q^2)$ and $F_{5N}(x,Q^2)$, and the suppression decreases with the increase in energy. For example, 
    at $E=10$ GeV the difference between the results obtained without and with $F_{4N}(x,Q^2)$ and $F_{5N}(x,Q^2)$ is $40\%(64\%)$ which becomes $23\%(37\%)$ and $17\%(25\%)$ at $E=30$ GeV and $E=50$ GeV, respectively for $\nu_\tau(\bar\nu_\tau)-N$ DIS process when a CoM energy cut of 1.4 GeV is used in the numerical calculations. When we increase the CoM energy cut to 2 GeV, it has been observed that the difference between the results obtained without and with $F_{4N}(x,Q^2)$ and $F_{5N}(x,Q^2)$ is $44\%(76\%)$ which becomes $24\%(43\%)$ and $17\%(33\%)$ at $E=30$ GeV and $E=50$ GeV, respectively for the $\nu_\tau(\bar\nu_\tau)$ scattering on the nucleon targets. Thus the increase in the value of CoM energy cut results in a larger reduction in the scattering cross section, specially in the case of antineutrino induced scattering.

 In Fig.~\ref{fig:sig_comp}, we have compared the results for the total scattering cross section $\sigma/E$ vs $E$ with the results of the different models available in the literature like that of Pashchos et al. (dashed line with diamond), Kretzer et al. (solid line with right triangle without a cut on $W$; dotted line with a cut of $W > 1.4$ GeV), Jeong et al. (dash-dotted line), Gazizov et al. (solid line with down triangle), Hagiwara et al. (solid line with cross symbol), Anelli et al. (double dash-dotted line) and Li et al. (solid line with circles) ~\cite{Kretzer:2002fr, Jeong:2010nt, Hagiwara:2003di, Conrad:2010mh, Paschos:2001np, Gazizov:2016, Li:2017dbe, Anelli:2015pba} as well as with the Monte Carlo generator GENIE~\cite{GENIE} and NuWro~\cite{NuWro}. The results are presented for both cases of cuts taken to be 1.4 GeV(shown by dashed line) and 2 GeV(shown by the solid line) by incorporating the target mass correction and higher twist effects at NLO in the four flavor scheme. 
 Our results with a cut of $W>1.4~ GeV$ (shown by dashed line) is in good agreement with the result of Kretzer et al.~\cite{Kretzer:2002fr} (shown by the dotted line) while there are significant differences from the result of Jeong et al.~\cite{Jeong:2010nt} (shown by the dash-dotted line). Notice that the results of the total scattering cross section with the same CoM energy cut reported by Kretzer and Reno~\cite{Kretzer:2002fr} and Jeong and Reno~\cite{Jeong:2010nt} are  also different with each other. The difference is mainly due to the choice of lower cuts on $Q^2$ in the evaluation of PDFs. It is important to point out that the results given by the different models~\cite{Kretzer:2002fr, Jeong:2010nt, Hagiwara:2003di, Conrad:2010mh, Paschos:2001np, Gazizov:2016, Li:2017dbe, Anelli:2015pba} have significant differences due to their choice of different kinematic regions.  Furthermore, we have observed that the effect of CoM energy cut is more pronounced in the case of $\bar\nu_\tau-N$ DIS than in $\nu_\tau-N$ DIS process. 
  Moreover, one may also notice that the total scattering cross section gets suppressed with the increase in the kinematic cut on the  CoM energy. It implies that a suitable choice of CoM energy cut ($W$) as well as four momentum transfer square ($Q^2$) to define the deep inelastic region and using them to calculate the nucleon structure functions, differential and total 
  scattering cross sections is quite important. These kinematic values should be kept in mind while comparing the predictions of 
  the cross sections in various theoretical models.

 In Fig.~\ref{fig:sig_ratio}, the ratio of total scattering cross sections, i.e. $\frac{\sigma_{\nu_\tau - N}}{\sigma_{\nu_\mu - N}}$ vs $E$ (dashed and solid lines) and $\frac{\sigma_{\bar{\nu}_\tau - N}}{\sigma_{\bar{\nu}_\mu - N}}$ vs $E$ 
  (dash-dotted and double dash-dotted lines) with a cut of $W>1.4$ GeV and $W>2$ GeV are shown. These results are evaluated at NLO with  the target mass correction effect in the three flavor scheme.
  These ratios show the effect of lepton mass in the total scattering cross section. Notice that the lepton mass effect is important through out the energy region shown here. However, this effect becomes small with the increase in energy and therefore the ratio increases but does not reaches unity even at 100 GeV. It is important to point out that for the ratio with CoM energy cut of 2 GeV, lepton mass effect is more pronounced than in the case of $W>1.4$ GeV cut on CoM energy. One may also notice that the lepton mass effect is quantitatively 
  different for neutrino and antineutrino induced processes, though qualitatively it shows a similar behavior. For example, the ratio obatined with a cut of $W>2$ GeV deviates from unity by $89\%(36\%)$ for neutrino and $91\%(38\%)$ for antineutrino at $E=10(50)$ GeV.

\section{Summary and conclusions}\label{summary}

In this work we presented the results for the nucleon structure functions, the differential and the total scattering cross sections for $\nu_\tau/\bar\nu_\tau-N$ DIS including TMC and HT effects and compared them with the predictions of the earlier calculations available in the literature and our findings are as follows: 

\begin{itemize}
\item The inclusion of perturbative and nonperturbative effects is quite important in the evaluation of nucleon structure functions, the differential and the total scattering cross sections. These effects are both $x$ and $Q^2$ dependent i.e. they are effective in the
 different regions of $x$ and $Q^2$.
\item The difference in the results of free nucleon structure functions $F_{iN} (x, Q^2);~~(i = 1-5)$ evaluated at NLO with and without the TMC effect is non-negligible. In the case of $F_{4N}(x,Q^2)$ this difference is quite large and it comes due to the TMC effect at mid and high $x$. When HT effect is also included in the evaluation of $F_{iN} (x, Q^2);~~ (i = 1-3)$, then we find that there is small difference in $F_{1N} (x, Q^2)$ and $F_{2N} (x, Q^2)$. The difference is slightly higher in $F_{3N} (x, Q^2)$ structure functions. With the increase in $x$ the effect of HT increases. We find that the difference due to HT effect is somewhat larger for $F_{3N}(x,Q^2)$ at low $x$ and low $Q^2$ which becomes small with the increase in $Q^2$. 
We conclude that with the increase in $Q^2$, both the TMC and HT effects become small in $F_{iN} (x, Q^2);~~ (i = 1-5)$ but are non-negligible for intermediate and high $x$.

\item The results for the differential scattering cross section with the inclusion of $F_{4N}(x, Q^2)$ and $F_{5N}(x, Q^2)$ structure functions leads to a large reduction in the cross section for $\nu_\tau$ and ${\bar\nu}_\tau$ scattering on the nucleons 
specially in the peak region. We have found that at lower energies, in the region of intermediate and high $y$ the differential scattering cross section for $\nu_\tau (\bar\nu_\tau)-N$ DIS becomes almost negligible unlike the case of $\nu_\mu (\bar\nu_\mu)-N$ DIS process. Furthermore, the contribution of $F_{4N}(x, Q^2)$ and $F_{5N}(x, Q^2)$ to the total scattering cross section is found to be more pronounced with the increase in (anti)neutrino energy. Quantitatively, this reduction depends upon the CoM energy cut ($W$) and the effect of increase in the cut on $W$ results in a significant decrease in the cross section as well as the peak shifts towards higher $y$.

\item  The effect of tau lepton mass results in a large reduction in the differential as well as the total scattering cross sections.  We explicitly show this reduction by numerically evaluating $\frac{d\sigma}{dy}$ and ${\sigma}$ for $\nu_\mu$ and $\nu_\tau$ induced charged current lepton production processes on the free nucleon targets. With the increase in neutrino energies this difference becomes gradually smaller but still the $\nu_\tau-N$ cross section is only 60-65$\%$ of $\nu_\mu-N$ cross section even at 50 GeV of neutrino energies, while it becomes 80-85$\%$ at 100 GeV.

 To conclude, the various perturbative and nonperturbative effects considered in this paper are effective in the various regions of $x$ and $Q^2$ and quite important in the energy region of 
 $E_{\nu_\tau/{\bar\nu}_\tau}~~<~~15$ GeV. The present theoretical results for $\nu_\tau/\bar\nu_\tau-N$ DIS scattering can be applied to obtain $\nu_\tau/\bar\nu_\tau -A$ DIS scattering cross sections in the nuclear targets like $^{16}O$, $^{40}Ar$, $^{208}Pb$ for which the work is in progress.
\end{itemize}

\section*{Acknowledgment}   

We are thankful to J. Sobczyk for providing DIS cross sections being used in NuWro MC and Steven Gardiner for providing DIS cross sections being used in GENIE. F. Zaidi is thankful to the Council of Scientific \& Industrial Research (CSIR), India, for providing the research associate fellowship with 
award letter no. 09/112(0622)2K19 EMR-I.
M. S. A. is thankful to the Department of Science and Technology (DST), Government of India for providing 
financial assistance under Grant No. SR/MF/PS-01/2016-AMU/G.

\end{document}